\documentclass[apj]{emulateapj}
\usepackage{apjfonts,amsmath}

\shorttitle{CO($1\rightarrow0$) Emission from SMGs}
\shortauthors{Hainline et al.}

\newcommand{\smmD}{SMM\,13120}
\newcommand{\smmN}{SMM\,09431}
\newcommand{\msun}{M_{\sun}}
\newcommand{\lsun}{L_{\sun}}
\newcommand{\kms}{\textrm{km\,s}^{-1}}
\newcommand{\jykms}{\textrm{Jy\,km\,s}^{-1}}
\newcommand{\hh}{\textrm{H}_{2}}
\newcommand{\mpc}{\textrm{Mpc}}
\newcommand{\kpc}{\textrm{kpc}}
\newcommand{\Kkmspc}{\textrm{K}\,\textrm{km\,s}^{-1}\,\textrm{pc}^{2}}

\begin{document}

\title{Observing Cold Gas in Submillimeter Galaxies: Detection 
of CO($1\rightarrow0$) Emission in \mbox{SMM J13120+4242} with the Green Bank Telescope}

\author{Laura J. Hainline\altaffilmark{1}, A. W. Blain\altaffilmark{1},
 T. R. Greve\altaffilmark{1}, S. C. Chapman\altaffilmark{1},
Ian Smail\altaffilmark{2}, and R. J. Ivison\altaffilmark{3,4}}
\altaffiltext{1}{Dept.\ of Astronomy, California Institute
of Technology, Mail Code 105-24, Pasadena, CA 91125, USA,
ljh@astro.caltech.edu, awb@astro.caltech.edu, tgreve@submm.caltech.edu, 
schapman@submm.caltech.edu}
\altaffiltext{2}{Institute for Computational Cosmology, Durham University,
South Road, Durham DH1 3LE, UK, ian.smail@durham.ac.uk}
\altaffiltext{3}{Institute for Astronomy, University of Edinburgh, Blackford Hill,
Edinburgh EH9 3HJ, UK, rji@roe.ac.uk}
\altaffiltext{4}{UK Astronomy Technology Centre, Royal Observatory, Blackford Hill,
Edinburgh EH9 3HJ, UK}

\begin{abstract}

We report the first detection of CO(1$\rightarrow$0) emission from a
submillimeter-selected galaxy, using the Green Bank Telescope.  We
identify the line in the spectrum of SMM\,J13120+4242 
as a broad emission feature at $z=3.408$, with $\Delta V_{FWHM}= 1040 \pm 190\,\kms$.
If the observed CO(1$\rightarrow$0) line profile arises from a single object 
and not several merging objects, then 
the CO(4$\rightarrow$3)/CO(1$\rightarrow$0) brightness temperature
ratio of $\sim 0.26$ suggests $n(\hh) > 3-10\times 10^{2}\,\textrm{cm}^{-3}$ and
the presence of sub-thermally excited gas. The integrated 
line flux implies a cold molecular gas mass $M(\hh)=1.6 \times 10^{11}\,\msun$, 
comparable to the dynamical mass estimate
and 4 times larger than the $\textrm{H}_{2}$ mass predicted from the
CO(4$\rightarrow$3) line assuming a brightness temperature ratio
of 1.0.  While our observations confirm that this
submillimeter galaxy is massive and gas-rich, they also suggest 
that extrapolating gas masses from $J_{upper}\geq 3$ transitions
of CO leads to considerable uncertainties.  
We also report an upper limit to the mass of cold
molecular gas in a second submillimeter galaxy, SMM\,J09431+4700, of 
$M(\hh)\lesssim 4\times 10^{10}\,\msun$.

\end{abstract}

 
\section{INTRODUCTION}

The resolution of the far-infrared (far-IR) 
background at long wavelengths through ground-based submillimeter-
and millimeter-wave surveys has revealed a
population of infrared-luminous, but optically faint, high-redshift galaxies,
whose spectral energy distributions (SEDs) and submillimeter flux densities suggest
large bolometric luminosities ($\sim 10^{13}\,\lsun$), mostly emitted in 
infrared (IR) wavebands.  Subsequent deep X-ray
observations by \citet{alexander03,alexander05}\ indicate that 
the bulk of this emission is not powered by active galactic nuclei (AGN), 
suggesting that high rates of star formation ($\sim 10^{3}\,\msun\,\textrm{yr}^{-1}$)
are responsible for the majority of the enormous luminosities of 
these submillimeter-selected galaxies (SMGs), even though 
they appear to already possess significant stellar masses ($\sim 10^{11}\,\msun$;
Borys et al.\ 2005).
SMGs have a surface density on the sky a factor of 10 smaller than 
the optically-selected, star forming, Lyman break galaxies, but
are 10 times as luminous; this, together with the fact that SMGs
seem to be strongly clustered \citep{blain04cluster}, would suggest that 
SMGs are an important population of forming galaxies, and either represent
the formation epoch of some of the most massive galaxies or trace
out unusually active regions.  However, 
the role they play in the scenario of hierarchical galaxy 
formation and evolution is not yet fully understood.  The large
far-IR luminosity and significant optical obscuration of SMGs \citep{smail04}\
are reminiscent of ultraluminous IR galaxies (ULIRGs) observed in
the local universe, most of which are strongly interacting or merging
systems \citep{ulirg96review}.  Since ULIRGs also show evidence for evolving 
into elliptical galaxies \citep{genzel01,tacconi02}, it has
been suggested that SMGs are high-$z$ analogs of ULIRGs, possibly progenitors
of the large bulges or elliptical galaxies observed locally 
\citep{smail02}.  However, determining
the baryonic and dynamical masses of SMGs, their system dynamics,
the duration of the phenomenon, and their end state, 
and matching them with theoretical predictions as well as with local
objects has proven a major challenge ---  
the faintness of SMGs at optical and near-IR wavelengths precludes
the detailed study needed to assess commonly-used evolutionary 
indicators.  

Studies of molecular gas emission provide the crucial mass and dynamical
information needed to evaluate the evolution and mass of SMGs
in the context of hierarchical galaxy formation and evolution.
The total intensity of molecular emission indicates the mass of gas
that is available to fuel future star formation, and the brightness 
temperature ratios from rotational transitions of different 
quantum number $J$ constrain
the temperature and density of the gas, suggesting how the gas is being consumed.
Even though millimeter-wave CO emission from SMGs is near currently achievable detection
limits, a series of interferometric surveys 
\citep[e.g.][]{frayer98,frayer99,neri03,grevesmg,tacconi06}\
have successfully detected $J\geq3$ CO($J\rightarrow J-1$) transitions
in SMGs, finding that SMGs are massive 
\citep[median $M_{dyn}>7\times10^{10}\,\msun$;][]{tacconi06},
gas-rich \citep[median $M(\hh)=3.0\times10^{10}\,\msun$;][]{grevesmg},
and compact \citep[median $D< 4\,\kpc$;][]{tacconi06}, consistent
with the hypotheses that SMGs are high-$z$ counterparts of ULIRGs and progenitors
of massive spheroids or ellipticals.
All of this information derives from warm molecular gas,
but we have limited knowledge of the less excited, cold molecular gas that
makes up a substantial fraction by mass
of the gas content of present-day galaxies (e.g.\ the observations of M82
of Wei\ss, Walter, \& Scoville 2005b). 
With a low threshold for excitation,
requiring $n(\hh) \sim 10^{2}\,\textrm{cm}^{-3}$ and $\Delta E/k \sim 5$\,K,
CO(1$\rightarrow$0) emission ($\nu_{rest}=115.271$\,GHz) is the most
representative tracer of the metal-enriched molecular gas mass
in galaxies because it is sensitive to the cold, diffuse gas
that may dominate.  \citet{papadop02}\ and Greve, Ivison, \& Papadopoulos (2003)
present evidence for large amounts of cold, low-excitation gas at high redshift, 
as first suggested by \citet{apm_co10}, in the galaxy HR~10.  
In this submillimeter-bright ERO that
lacks obvious AGN signatures, at $z_{CO}=1.439$, the CO(5$\rightarrow$4) line
luminosity underestimates the cold molecular
gas mass by nearly an order of magnitude.
Observations of CO($1\rightarrow0$) in SMGs 
are thus complementary to studies of $J\geq3$ gas and necessary to determine
the total mass of molecular gas they contain.

Previous studies of CO(1$\rightarrow$0) emission from high-$z$
galaxies have been carried out with the Very Large Array
(VLA) (e.g.\ Papadopoulos et al.\ 2000; Carilli et al.\ 2002b; 
Greve et al.\ 2003; Greve, Ivison, \& Papadopoulos 2004).
The Robert C.\ Byrd Green Bank Telescope (GBT) far
exceeds the current capabilities of the VLA in
instantaneous spectral bandwidth (800\,MHz maximum per spectrometer bank, compared 
with 50\,MHz at the VLA) with a comparable effective collecting area.  
Since high-$z$ CO lines can be wider than the VLA
bandwidth \citep{thomasRG}, the GBT is better suited to search for and measure
accurately the spectral shape of CO emission from SMGs, at the
expense of limited spatial resolution. 

In this paper, we present the results of a search
for CO(1$\rightarrow$0) emission from SMGs, undertaken with the GBT
and the facility's K-band receiver (18--26\,GHz). 
The tuning range of the GBT's K-band receiver restricts
our potential SMG targets to those with redshifts greater than $z\simeq 3.35$, of which
there are only four with measured redshifts \citep{chapman05,ledlow02}.  
Of these four SMGs, we chose to target 
for GBT observations the two which also have previous high-$J$ CO
detections (both from CO(4$\rightarrow$3) observations at the
IRAM Plateau de Bure interferometer) so we could 
compare the line strengths and widths of any detections: 
SMM\,J09431+4700 at $z_{CO}=3.346$ (Neri et al.\ 2003; hereafter \smmN)
and SMM\,J13120+4242 at $z_{CO}=3.408$ (Greve et al.\ 2005; hereafter \smmD).
Throughout our discussion, we assume $\Omega_{\textrm{M}}=0.3$, 
$\Omega_{\Lambda}=0.7$, and $H_{0}=70~\kms~\mpc^{-1}$. 

\section{OBSERVATIONS}

We searched for CO(1$\rightarrow$0) emission from \smmN \ and
\smmD \ with the 18--26\,GHz dual-feed, dual-polarization 
receiver (K-band) at the GBT on UT 2004 December 1. We utilized the 
position-switching ``Nod'' observing pattern, in which the two feeds
alternate between the target position and a blank-sky position $3\arcmin$
away in azimuth such that one feed (beam) is always on the target source.  
A 2-minute integration (scan) was completed before
the receiver moved in azimuth to the alternate position; thus a full Nod sequence,
consisting of a pair of scans, lasted 4 minutes. Every 90 minutes, the
pointing and focus of the telescope were checked on a nearby bright radio 
source. Observing conditions were excellent through both night and day,
with low atmospheric opacity ($\simeq 0.02$\,nepers) and system temperatures
in the range 30--50\,K.  Total on-source integration times
were 6.0 hours for \smmN \ and 6.2 hours for \smmD.
The size of the main beam at the GBT at 26\,GHz is $28\arcsec$, so
our observations are not spatially resolved.  However, since $1\arcsec$ 
corresponds to $\sim 7.4$\,kpc at $z\simeq3.4$, it is very likely 
the entire extent of our target galaxies and any companions will be within 
the main beam ($\sim 210$\,kpc at $z\simeq3.4$) and our observations should be
representative of the average cold gas properties over the observed galaxy.

We used the GBT's Autocorrelation Spectrometer (ACS) in 3-level sampling mode.
For each of the two target sources, we observed a 800\,MHz bandpass 
centered on the frequency at which the CO(1$\rightarrow$0) line 
should fall based on the CO system redshift obtained from
the previous CO(4$\rightarrow$3) detection.
The spectrometer banks were configured in low-resolution mode, providing 
2048 channels of width 391\,kHz each ($\sim 4.4\,\kms$ at 26\,GHz) for both 
of the two circular polarizations of each feed.  The noise diode for the
K-band receiver was fired every 1.0\,sec for calibration;
we average the data taken with the calibration noise diode on with
that taken with the noise diode off to improve our signal-to-noise (S/N) 
ratio.  The noise diode temperature ($T_{cal}$) values for the K-band
receiver provided by the observatory were verified using
spectral observations of the radio-loud quasars 3C\,147 
(for \smmN) and 3C\,295 (for \smmD) taken with the same 
spectrometer setup as the science targets.

\section{CALIBRATION}

In single-dish radio observations, the spectral baseline is the 
limiting factor in detecting very faint, broad lines like those 
expected to be produced by CO(1$\rightarrow$0) emission
from a high-$z$ galaxy.  Vanden Bout, Solomon, \& Maddalena (2004) discuss the significant 
baseline curvature seen in wide-bandwidth GBT spectral data obtained
over a range of elevations and timescales,
despite the off-axis feed arm design of the telescope, which
was intended to reduce reflections of signal along the optical path.
The spectral baselines at the GBT are caused by a combination of small differences
in the telescope system power between on- and off-source scans and 
reflections between components in the optical system, and cannot
be simply modeled with a low-order ($n<10$) polynomial.  Our GBT data are
certainly affected by complex baseline shapes, since we average
over $\sim 6$\,hours of integration and $60\degr$ of elevation. The calibration
of our spectral data is thus nontrivial; we 
must take special care to minimize the systematic errors
in our spectra associated with baseline ripples or curvature, and strive 
to not introduce any line-scale structure into our spectrum while attempting
to remove the baseline.

\Citet{vsmHCN}\ outline a method for calibrating GBT spectral data of
faint sources, in which the baseline shape is determined directly from the data
and subtracted.
They first produce a difference spectrum for both polarizations of each
beam for every pair of scans that comprise a full Nod sequence according 
to the formula
\begin{equation}
T_{diff}(\nu)=\frac{ON - OFF}{OFF}\,T_{sys},
\end{equation}
where $ON$ and $OFF$ represent an on-source scan and a blank-sky reference scan,
respectively, observed with the same receiver feed.  The resulting spectra are
averaged over time and feed, weighted by $(T_{sys})^{-2}$, to
produce a master average science spectrum $T(\nu)$ for each polarization.
The template containing the baseline shape of each master spectrum is constructed next
by first taking the scans in pairs of Nod sequences (four scans) and subtracting 
successive ON scans and OFF scans in the following manner:
\begin{subequations}
\begin{eqnarray}
\frac{(ON_a - ON_b)\,_{beam 1}}{(ON_b)\,_{beam 1}}\,T_{sys,\,b}, \\
\frac{(OFF_a - OFF_b)\,_{beam 1}}{(OFF_b)\,_{beam 1}}\,T_{sys,\,b}, \\
\frac{(ON_a - ON_b)\,_{beam 2}}{(ON_b)\,_{beam 2}}\,T_{sys,\,b}, \\
\frac{(OFF_a - OFF_b)\,_{beam 2}}{(OFF_b)\,_{beam 2}}\,T_{sys,\,b},
\end{eqnarray}
\end{subequations}
where $ON_a$ and $ON_{b}$ are the ON scans from the first and
second Nod sequences, respectively, and similarly for $OFF_a$ and $OFF_b$.
Then, the sequence of ON-ON and OFF-OFF spectra is averaged to
form the master baseline template $B(\nu)$ for each polarization.  
\Citet{vsmHCN}\ point out that 
$B(\nu)$ contains the same baseline structure as the science spectrum 
$T(\nu)$ (though with higher amplitude due to the
longer time delay between differenced quantities), but without signal from
the emission line being observed.  Hence, to remove the baseline
described by the template so that the only spectral
shape remaining is due to the observed source, $B(\nu)$ 
is smoothed and fit to the master science spectrum for each polarization
by finding the best-fit coefficients $(a,b,c)$ such that
\begin{equation}
T(\nu)= a B(\nu) + b\,\nu + c.
\end{equation}
The best-fit template is then subtracted from $T(\nu)$. 
In Figure~1(a) we show the final averaged spectrum after applying this method of
baseline reconstruction and subtraction to our Feed~1 data on \smmN \ and 
averaging the two polarizations of the feed together. 
The majority of the baseline shape has clearly been removed
in Figure~1(a), but a slight residual baseline remains.
While this residual is less complex than the original baseline
shape, its origin is unclear; this causes it to be difficult to 
fit and subtract, which may interfere with identification of 
faint lines from the target source.
\begin{figure}
\includegraphics[width=\columnwidth]{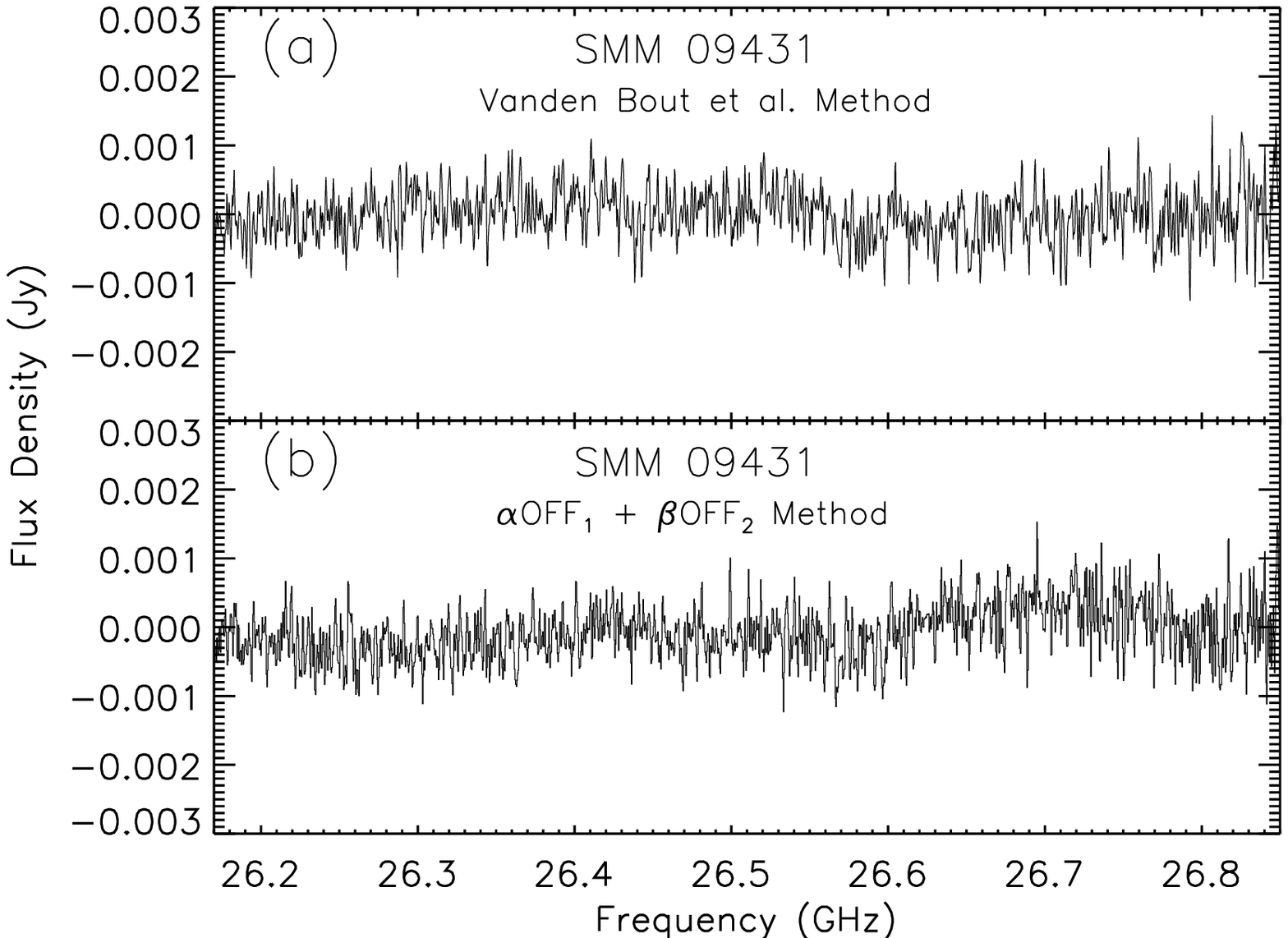}
\caption{(a) Final averaged spectrum of Feed~1 for \smmN \ data, 
produced by the GBT spectral calibration 
method of \citet{vsmHCN}.  The RMS per spectral channel is 0.37\,mJy.
(b) Final averaged spectrum resulting from the application of the
$\alpha$OFF$_{1}$ + $\beta$OFF$_{2}$ calibration method on the same 
data.  The RMS per spectral channel is 0.41\,mJy.  
Based on the CO(4$\rightarrow$3) redshift of
$z=3.346$ from \citet{neri03}, we expect to see the CO(1$\rightarrow$0)
line at $\nu_{obs}=26.523$\,GHz.}
\end{figure}
 
We preferred not to fit and subtract a baseline template from 
our final averaged spectrum at all, hoping to avoid the
uncertainties associated with the process.  As a result, we 
chose to use a different method of calibration, a variation of the traditional method
of single-dish radio spectrum calibration described in Equation~1,
which we refer to as the $\alpha$OFF$_{1}$ + $\beta$OFF$_{2}$ method.
In effect, we use the two OFF scans surrounding each ON scan 
to construct an improved, interpolated OFF, separately for each 
polarization in each feed. However, instead of assuming that
the preceding and following OFFs contribute equally
to the spectral shape of each ON scan, which might be expected
if the spectral baseline was purely a linear function of time, we use
least-squares fitting to find values $\alpha$ and $\beta$ 
for each scan such that
\begin{equation}
\frac{ON - (\alpha\,OFF_{1} + \beta\,OFF_{2})}{
\alpha\,OFF_{1} + \beta\,OFF_{2} } = 0,
\end{equation}
where $OFF_{1}$ refers to the scan prior to the ON scan and
$OFF_{2}$ refers to the scan following the ON scan (all 
taken through the same feed). Allowing $\alpha$ and $\beta$ to vary by scan
helps remove systematic shapes caused by sporadic phenomena
of varying amplitude; we found such coefficient variation necessary
because even though for $\sim 70\%$ of scans the ($\alpha$, $\beta$) values were 
($0.5\pm0.05$,$0.5\pm0.05$), it was not unusual (the remaining $\sim 30\%$ of scans) 
for $\alpha$ and $\beta$ to lie in the range 0.25--0.45 or 0.55--0.75, where
$\beta = 1 - \alpha$.
After the spectral fits, we apply the derived values of $\alpha$ and
$\beta$ to obtain the normalized difference spectrum, for each scan,
in units of antenna temperature:
\begin{equation}
T_{diff}(\nu) = \frac{ON - (\alpha\,OFF_{1} + 
\beta\,OFF_{2})}{\alpha\,OFF_{1} + 
\beta\,OFF_{2} } (\alpha\,T_{sys1} + \beta\,T_{sys2}),
\end{equation}
where $T_{sys1}$ is the system temperature derived for the first OFF scan 
and $T_{sys2}$ is the system temperature derived for the
second OFF scan.  The difference spectra from all scans are then averaged together, 
weighted by the root-mean-square deviation (RMS) of each scan.  
In Figure 1(b) we show the result of the $\alpha$OFF$_{1}$ + $\beta$OFF$_{2}$ method
on the same spectral data used in Figure 1(a).  The final averaged spectrum in Figure~1(b)
lacks obvious residual ripples on the scale of a galaxy emission line
($\sim 100-1000\,\kms$, or $\sim 9-90$\,MHz at $\nu_{obs}=26.5$\,GHz), though a long
period ($\sim$ few hundred MHz), low amplitude ripple may be present which is not likely
to hinder faint line detection.  Note that this reasonably flat spectrum 
was produced without subtracting a curved or rippled baseline template. 
In the case of \smmD, a clearly linear baseline 
appeared in the final averaged spectrum produced by this
method, which was easily fit and removed.  
In general, we find that the $\alpha$OFF$_{1}$ + $\beta$OFF$_{2}$ method
produces qualitatively similar results to that
used by \citet{vsmHCN}: sources detected with one calibration
method are also detected by the other method, and the resulting
spectra have comparable single-channel RMS values (the RMS values of the
final average spectra, in the sense 
vanden Bout et al.\ method versus $\alpha$OFF$_{1}$ + $\beta$OFF$_{2}$ method,
are 0.37\,mJy versus 0.41\,mJy for \smmN \ and 
0.44\,mJy versus 0.42\,mJy for \smmD).  In addition, neither
method can completely remove all baseline shape from our GBT data.
Consequently, we cannot say that one method produces better results
than the other.  Even so, we prefer the 
$\alpha$OFF$_{1}$ + $\beta$OFF$_{2}$ method because no curved baseline
subtraction is required.  However, it is important to note that the
method is suitable only for sources with very
weak or undetectable continuum emission at $\lambda_{rest} = 2.6$\,mm.
SEDs of SMGs \citep[e.g.][]{chapman05}\ predict that continuum emission from SMGs
at 30\,GHz is undetectable in 6\,hours of integration time with the GBT.  

Before averaging the spectral data from the different polarizations and 
feeds, we examined the periodograms (Fourier power spectra) of the calibrated spectra from
both polarizations and feeds of the individual scans, looking for any
peaks that might indicate systematic spectral features 
not removed in the calibration process.  Such ripples could influence 
the appearance of the spectrum, and we were especially concerned about 
any ripples of width similar to the expected width of a CO emission line 
from a high-$z$ source ($\sim 300\,\kms$, or $\sim 30$\,MHz). We found that
the right polarization of feed~2 nearly always showed an irregularly 
changing pattern of power between 0--0.03\,MHz$^{-1}$ (periods
$\gtrsim 30$\,MHz).  This feature was not found in feed~1
data, and was found in approximately half of the left polarization data
from feed~2.  Thus, in an effort to ensure that what appears as noise
on the scale of an emission line is due to random processes, we have not included  
any feed~2 data in the averaged calibrated spectra. While this exclusion
cuts our effective integration time in half and increases the spectral rms,  
it helps ensure that the quoted noise is uncorrelated in frequency, 
and increases our confidence in the reality of any possible emission 
line detection by removing ripples and structure
that could obscure the line. 

To convert our object spectra to units of Janskys, we corrected for
atmospheric attenuation and then applied the parametrization
of the GBT gain versus elevation curve found by \citet{condon03},
re-scaled for K-band by assuming an aperture efficiency at 26\,GHz of
0.55 from an interpolation of GBT aperture efficiency measurements at nearby
frequencies.  We estimate that our flux calibration 
is accurate to within 15\%, due to uncertainties in the temperatures of the 
calibrator diodes, the GBT gain curve, and the calibration noise diode variations
across the observed bandwidth, as well as the opacity estimate.  

All of the 
calibration of our spectral data was accomplished with the new
library of routines written by the GBT staff in IDL, GBTIDL\footnote{
Information on GBTIDL is available at http://gbtidl.sourceforge.net./.}.
  
\section{RESULTS}

\subsection{\smmD}

\smmD, also known as SSA\,13-332, was identified 
as a source in the Hawaii Deep Field SSA~13,
with $S_{850\micron}=6.2$\,mJy.  The galaxy, 
with an optical redshift of $z=3.405$, is not known to be lensed, 
and its optical spectrum reveals active nucleus signatures   
\citep[\ion{C}{4}, \ion{Si}{4}, \ion{O}{3};][]{chapman05}. 
\citet{chapman05}\ estimate from the radio-submillimeter SED
a bolometric luminosity for \smmD \ $L_{bol}=2.02\times 10^{13}\,\lsun$
and a characteristic dust temperature of $T_{d}=47$\,K, though
\citet{kovacs06}\ find from their $350\micron$ 
SHARC-II observations of SMGs from the \citet{chapman05}\ sample
that the luminosities presented in
\citet{chapman05}\ are typically overestimated by a factor of 2 and
the dust temperatures are overestimated by 13\%, which
would bring down the values for \smmD \ to 
$L_{bol}=1.01 \times 10^{13}\,\lsun$ and $T_{d} = 42$\,K.
High-$J$ CO observations have confirmed the redshift and 
shown \smmD \ to be very massive in molecular gas. 
\citet{grevesmg}\ detect CO(4$\rightarrow$3) emission 
from \smmD \ with the IRAM Plateau de Bure
interferometer at $z=3.408$, with velocity width $\Delta V_{FWHM}=530\pm50\,\kms$
and integrated line flux $S_{CO}\,\Delta V=1.7\pm0.3\,\jykms$.  
Assuming a CO luminosity-to-$\hh$ conversion factor 
$\alpha=0.8\,\msun\,(\Kkmspc)^{-1}$, this yields 
a molecular gas mass of $M(\hh)=(4.2\pm0.7)\times10^{10}\,\msun$.
\citet{grevesmg}\ also estimate a lower limit on the dynamical mass of
\smmD \ of $M_{dyn} \gtrsim 1.2\times10^{11}\,\msun$. 
\smmD \ is unresolved in the $6\farcs9 \times 4\farcs8$ beam at IRAM, so 
no information on the spatial structure of the molecular gas 
is available.  

We detect CO(1$\rightarrow$0) emission from \smmD \
centered at $\nu_{obs} = 26.1481$\,GHz, 
with an intensity-weighted line centroid in redshift space of $z=3.408\pm0.004$. 
This detection represents the first detection of 
CO(1$\rightarrow$0) from a \emph{bona fide} 
SMG and the first SMG CO(1$\rightarrow$0) detection with spectral information.  
In Figure~2 we show our calibrated 
and averaged spectrum of \smmD, produced by the $\alpha$OFF$_{1}$ + 
$\beta$OFF$_{2}$ method described in \S3.  We have subtracted a linear
baseline from the spectrum, fit to all the channels away from the line,
and smoothed it from its original velocity resolution of $4.47\,\kms$ and RMS per
spectral channel $\sigma=0.42$\,mJy to $94\,\kms$ and $\sigma=0.16$\,mJy.
The detected line is still visible if the data from feed~2
are included in the final average spectrum.
\begin{figure}
\includegraphics[width=\columnwidth]{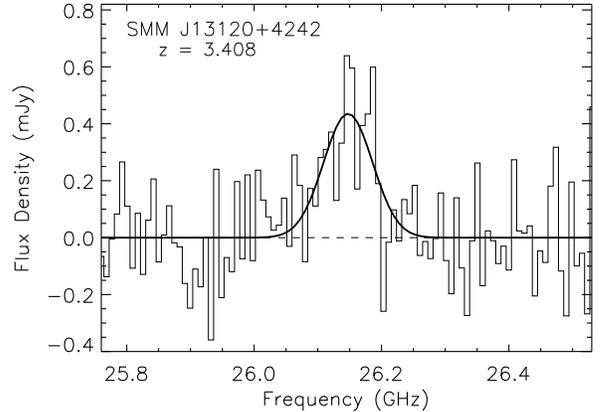}
\caption{CO(1$\rightarrow$0) spectrum of \smmD, produced by the
$\alpha$OFF$_{1}$ + $\beta$OFF$_{2}$ method.  A linear baseline
has been subtracted from the spectrum post-calibration. 
The spectrum has been boxcar-smoothed to a velocity 
resolution of 94\,$\kms$ and re-binned so that each channel
is independent.  A Gaussian fit to 
the line with FWHM $1040\,\kms$ is overplotted as the thick curve.}
\end{figure}

The Gaussian fit overplotted in Figure~2 shows a wide CO(1$\rightarrow$0) line,
$\Delta V_{FWHM}=1040\pm190\,\kms$.  However, the Gaussian fit to the line
profile shown in Figure~2 appears poor, possibly because of
the triple-peaked appearance of the line, although the $\chi^{2}$-value of
the Gaussian fit, $\chi^{2}=0.19$, indicates that we cannot 
formally conclude whether or not the Gaussian function 
provides a good description of the line profile. To obtain
a less model-dependent estimate of the line width,
we compute also the second moment of the emission spectrum, finding 
$\Delta V = 350\,\kms$.  Integrating over the
line, we find $S_{CO}\,\Delta V=0.42\pm0.07\,\jykms$, where the error
includes the (estimated) 15\% uncertainty in our flux calibration. 
Applying the relations for $L_{CO}$ and $L^{\prime}_{CO}$
presented in Solomon, Downes, \& Radford (1992), this
corresponds to a CO luminosity of $L_{CO} = (1.0 \pm 0.2) \times 10^{7}\,\lsun$ or 
$L^{\prime}_{CO}=(2.0 \pm 0.3) \times 10^{11}\,\Kkmspc$.
Adopting the CO luminosity-to-$\textrm{H}_{2}$ mass conversion factor
$\alpha=0.8\,M_{\sun}\,(\textrm{K\,km\,s}^{-1}\,\textrm{pc}^{2})^{-1}$,
which is found by \citet{downes98}\ to be appropriate for molecular
gas exposed to the intense ultraviolet (UV) radiation fields and
strong tidal forces present at the centers of nearby ULIRGs, 
we estimate the total mass of low-excitation
$\hh$ in \smmD \ to be $M(\textrm{H}_{2}) = (1.6 \pm 0.3) \times 10^{11}\,M_{\sun}$.
Note that the CO-to-H$_{2}$ mass conversion factor can
vary depending on local gas conditions \citep[e.g.][]{yao03}\ and is thus 
highly uncertain for galaxies at high redshift.  As a result, 
our absolute value of the molecular gas mass may have significant
systematic error that is difficult to quantify. Should the gas in
\smmD \ be less dense and less excited than that of ULIRGs, the 
conversion factor for \smmD \ should be larger and so we are 
underestimating the molecular gas mass; if the gas is more 
dense and more excited, the conversion factor may be too high,
and we are overestimating the gas mass.

\subsection{\smmN}

\smmN \ was identified by Cowie, Barger, \& Kneib (2002) in 
deep submillimeter maps of the $z=0.41$ galaxy cluster Abell~851 obtained 
with the SCUBA instrument on the James Clerk Maxwell Telescope.  
From subsequent 1.4~GHz observations, \citet{ledlow02}\ estimated $T_{d}=38\,$K,
and suggested that two individual radio sources, identified as H6 and H7, 
were associated with the SCUBA source.  Optical follow-up
observations showed H6 to be a narrow-line Seyfert 1 galaxy 
at $z=3.349$ \citep{ledlow02}, while the weaker radio source H7
has a redshift of $z=3.347$ from \ion{O}{3}\
(Takata et al. 2006, in preparation).  Both \citet{neri03}\ and \citet{tacconi06}\
detect CO(4$\rightarrow$3) emission from H7 at $z=3.3460\pm 0.0001$ with 
a line width $\Delta V_{FWHM} = 420\pm 50\,\kms$ and 
an integrated line flux $S_{CO}\,\Delta V = 1.1\pm 0.1\,\jykms$.  
H6 has not been definitively detected in CO. Assuming a CO to $\hh$ 
conversion factor $\alpha=0.8\,\msun~(\Kkmspc)^{-1}$, 
\citet{neri03}\ infer a molecular gas mass $M(\hh)= (2.1\pm 0.2) \times10^{10}\,\msun$.  
\smmN \ is lensed by the foreground galaxy cluster by a modest factor of 1.2 
\citep{cowie02}; this factor has been included in all stated values of
mass and luminosity.

We do not detect CO(1$\rightarrow$0) emission from \smmN, which is
apparent in the object's final calibrated spectrum displayed in Figure~1(b).
Our observations yield a single-channel RMS of 0.41\,mJy,
which in turn corresponds to a statistical $3\sigma$ upper limit to the
integrated flux density at 26.523\,GHz of 0.056\,$\jykms$, assuming
the CO(1$\rightarrow$0) line follows the CO(4$\rightarrow$3) line 
FWHM of 420\,$\kms$ and that the line has a Gaussian profile. 
However, due to the uncertainties associated with the GBT spectral 
baselines and the flux calibration, we feel that the value of the
error in the integrated flux,
$\sigma_{S\,\Delta V}$, found for \smmN \ through application of the statistical
error formula is probably an underestimate and thus will result in
unrealistically low upper limits on the CO line luminosity and
molecular gas mass.  Consequently, we estimate an alternate
value of $\sigma_{S\,\Delta V}$ for \smmN \ from the error in
the integrated flux of \smmD, noting that the single-channel RMS values
of our final spectra of both objects are nearly identical 
(0.41\,mJy for \smmN, 0.42\,mJy for \smmD) and that the objects
were observed in similar conditions on the same day at nearly the same 
sky frequency.  We scale the error in $S_{CO}\,\Delta V$ of \smmD \ 
($\sigma_{S\,\Delta V}=0.07\,\jykms$, see \S4.1) by 
$[\Delta V_{FWHM}(\textrm{\smmN})/\Delta V_{FWHM}(\textrm{\smmD})]^{1/2}$
to account for the difference in CO line width between the two objects,
and thus obtain the more conservative $3\sigma$ upper limit of
$S_{CO}\,\Delta V < 0.13\,\jykms$ for \smmN.  For the sake of 
completeness, we provide the 
$3\sigma$ upper limits of CO line luminosity and $M(\hh)$ that result
from both the statistical upper limit on the integrated line flux and 
our alternative upper limit.  From $S_{CO}\,\Delta V < 0.056\,\jykms$,
we obtain the upper limits to the CO luminosity 
of \smmN \ of $L_{CO}<1.1\times10^{6}\,\lsun$, or
$L^{\prime}_{CO}<2.2 \times10^{10}\,\Kkmspc$.  Assuming the CO-to-$\hh$
conversion factor $\alpha=0.8\,M_{\sun}\,(\textrm{K\,km\,s}^{-1}\,\textrm{pc}^{2})^{-1}$,
the $3\sigma$ upper limit on $M(\hh)$ for \smmN \ is then
$M(\textrm{H}_{2})<1.8\times10^{10}\,M_{\sun}$.  For the conservative estimate
$S_{CO}\,\Delta V < 0.13\,\jykms$, the corresponding upper
limits are $L_{CO}<2.5\times10^{6}\,\lsun$, $L^{\prime}_{CO}<5.1 \times10^{10}\,\Kkmspc$,
and $M(\textrm{H}_{2})<4.1\times10^{10}\,M_{\sun}$, which we believe
to be more reasonable than the statistical upper limits.  
If we were to assume a broader CO(1$\rightarrow$0) line, the upper limit will
be still larger and therefore less restrictive.  

\section{ANALYSIS \& DISCUSSION}

\subsection{CO Line Ratios and Gas Excitation}

With detections in multiple CO lines for \smmD, one of which is from
CO(1$\rightarrow$0), we can use the ratio of the brightness
temperatures of the different transitions to extract some 
clues about the globally-averaged conditions of its molecular gas.
The task is made easier because \smmD \ is not known to be 
strongly gravitationally lensed, so the effects of differential 
lensing on the line fluxes are not likely to complicate
our interpretation.  Defining $r_{43}$ as the ratio of the area/velocity averaged 
brightness temperature of the CO(4$\rightarrow$3) and 
CO(1$\rightarrow$0) lines, and assuming all of the
CO emission is from a single object, we find $r_{43}=0.26\pm0.06$ for \smmD.
This value of $r_{43}$ is notably lower than the ratio that 
is expected from a thermalized (i.e.\ relative level populations
described by principles of local thermodynamic equilibrium), optically
thick source, $r_{43,LTE}=1.0$, implying that the $J=4$ level of CO is not
thermalized in \smmD.  As a consequence, $M(\hh)$ estimates for 
\smmD \ calculated from $J_{upper}\geq4$ CO lines assuming the lines are
fully thermalized and optically thick will underestimate 
the mass of low-excitation gas in the galaxy.  This is exactly what
we find from our CO(1$\rightarrow$0) detection of \smmD: our estimate
of the molecular gas mass in \S4.1, 
$M(\hh) = 1.6 \times 10^{11}\,\msun$, is a
factor of $\sim 4$ greater than the mass implied by the CO(4$\rightarrow$3)
observations if constant brightness temperature
is assumed, $M(\hh)= 4.2 \times10^{10}\,\msun$.  Moreover,
the lack of a dominant population of thermalized high-excitation
gas in \smmD \ suggests that the CO luminosity-to-H$_2$ 
mass conversion factor we use in \S4.1 may be incorrect for this galaxy.  
\citet{downes98}\ derive that conversion factor for 
the central starbursts in ULIRGs, and 
from observations of the central IR-luminous starburst
region of M82 \citep[e.g.][]{weiss05a}\ we expect the CO
excitation in a starburst region to be thermalized at least to the 
CO $J=4$ level. Since the $J=4$ level is observed to not be thermalized when
averaging over \smmD, the CO-to-$\hh$ conversion factor
of \citet{downes98}\ is likely too low for \smmD, and the 
total molecular gas mass is likely to be even larger than the
$1.6\times10^{11}\,\msun$ we calculate in \S4.1. 
   
Such low values of $r_{43}$ as we find in \smmD \ are not unknown: 
in the Milky Way, $r_{43}$ ranges with position between $\sim 0.1$--0.2
(Fixsen, Bennett, \& Mather 1999).  The line ratio $r_{43}$ we find
for \smmD \ is also reminiscent of
the CO(5$\rightarrow$4)/CO(1$\rightarrow$0) ratio of 0.1 found in
the ERO/SMG HR~10 by \citet{HR10co10}, another high-$z$ object 
in which the high-$J$ CO emission does not trace the 
low-excitation gas. That \smmD \ and HR~10 are similar is somewhat
surprising, however, because HR~10 is not known to harbor an active nucleus, while
\smmD \ clearly does.  On the other hand, the ratio $r_{43}$ in \smmD \
is significantly smaller in comparison to those of the other
high-$z$ active galaxies with CO(1$\rightarrow$0) detections.
PSS~2322+1944, APM~08279+5255, and 4C\,60.07 are two QSOs and 
a radio galaxy, respectively, and have values of $r_{43}$ of 
1.4, 1.5, and 0.7, indicating that the molecular gas they contain
is nearly thermalized and that molecular gas masses derived 
from CO(4$\rightarrow$3) and CO(1$\rightarrow$0) emission will
be roughly similar, which is found to be true \citep{pss_co10,apm_co10,thomasRG}.
Since \smmD \ hosts an AGN, we might expect its gas 
conditions to be more like those of the high-$z$ active galaxies than
galaxies lacking active nuclei; however,
note that the active nucleus in \smmD \ is probably significantly less 
luminous than those found in PSS~2322+1944 and APM~08279+5255, 
because, in its rest-frame UV spectrum, the AGN lines are weak and
interstellar absorption features indicative of a starburst 
are also clearly visible \citep{chapman05}.
It is possible that the less-powerful active nucleus in \smmD \ could 
influence the gas properties of the host galaxy in a different
way than would a high-luminosity AGN, causing
the global CO line ratios to differ between the SMG and the QSOs.
It may thus be inappropriate to use the mere presence of an AGN
as a factor in predicting global CO line ratios.

From our measurement of $r_{43}$ in \smmD \ we can obtain 
quantitative constraints on the average kinetic
temperature and density of the molecular gas in this system.  To this end, we
assume the CO(1$\rightarrow$0) and (4$\rightarrow$3) emission
result from a single gas phase and use a large
velocity gradient (LVG) model \citep[based on][]{richardson85}\ 
to fit the observed line ratio.
Data from only two CO lines generally do not provide enough 
information to place strong constraints on the excitation; 
however, since one of the lines is from the
1$\rightarrow$0 transition, we can place strong lower limits on 
the column and number density of CO, and by extension, on 
the density of $\hh$. 
A standard LVG code, assuming 
a typical Galactic value for the [CO/H$_{2}$] abundance $X_{CO}=10^{-4}$,
velocity gradient $dV/dr=1\,\kms\,\textrm{pc}^{-1}$, and
$T_{CMB} = (2.73\,\textrm{K})(1+z)$, finds a best-fit density for \smmD \
of $n(\hh)>300~\textrm{cm}^{-3}$ for a range of gas kinetic
temperatures $T_{k}=42-54$\,K.  This solution is robust, such 
that the model finds no candidate solutions at other densities. 
However, in kinematically violent regions with 
intense UV radiation fields, such as we expect to find in ULIRGs
and SMGs, the [$\textrm{CO}/\hh$] abundance is likely to be
lower, so we also calculate the kinetic temperature and
density limits for $X_{CO}=10^{-5}$ to cover a range of 
possible values.  Leaving the assumed values of $dV/dr$ and
$T_{CMB}$ unchanged, we obtain a slightly higher limit
on the molecular gas density of $n(\hh)>10^{3}~\textrm{cm}^{-3}$
for a somewhat lower gas temperature range of $T_{k}=34-46$\,K.
Both of these sets of temperature and density limits point to
low-density gas in \smmD, though we would guess that the 
solution for $X_{CO}=10^{-5}$ is closer to reality. Also, 
both of the model fits result in kinetic temperatures that
are in good agreement with the dust temperature of \smmD \
found by \citet{chapman05}, corrected by a factor of 13\% 
as suggested by \citet{kovacs06}, that is, $T_{d}=42$\,K. 
$T_{k} \simeq T_{d}$ would be expected for an
interstellar medium in which the gas and dust are well mixed and
thermally coupled.  Encouraged by these similar values, 
we adopt a fixed value of
$T_{k}=T_{d}=42$\,K, leave $X_{CO}/(dv/dr)$ free to vary, 
and once more use the LVG model to calculate the density
of the molecular gas in \smmD, obtaining $n(\hh)>10^{3}~\textrm{cm}^{-3}$
and $X_{CO}/(dv/dr)=10^{-5}\,(\kms\,\textrm{pc}^{-1})^{-1}$, consistent
with the results of the previous fits. Therefore, the general
constraints we can place on the globally-averaged gas properties
of \smmD \ are $n(\hh)>(3-10)\times10^{2}\,\textrm{cm}^{-3}$
and $T_{k}=34-54$\,K, for a realistic range of
$X_{CO}/(dv/dr)=10^{-5}-10^{-4}\,(\kms\,\textrm{pc}^{-1})^{-1}$. The
mean gas density is thus likely to be relatively low in this system.    
Additionally, it is interesting to note that, if we assume the galaxy
is a uniform-density sphere, our density limits
imply a small upper limit to the radius of 
\smmD \ in the range 0.9--1.4\,kpc, which is
consistent with the median upper limit on SMG source size that 
\citet{tacconi06}\ derive from their high-resolution images
of high-$J$ CO emission from eight SMGs.

The diffuse, warm, and low excitation gas we find in \smmD \ 
has been observed in local ULIRGs \citep{solomon97}\
and at high redshift in HR~10 \citep{HR10co10}.   
In contrast, previous CO(1$\rightarrow$0) detections of other
$z>3$ active galaxies have implied lower limits on the cold gas density
nearly an order of magnitude larger. PSS~2322+1944, APM~08279+5255, 
and 4C\,60.07 have cold gas densities derived from $r_{43}$
and LVG models in the range $n(\hh)>10^{3}-10^{4}~\textrm{cm}^{-3}$  
\citep{pss_co10,apm_co10,thomasRG}.

For \smmN, we can use our $3\sigma$ upper limit calculations to
place only a lower limit on $r_{43}$.  We find $r_{43}> 1.2$ for
$S_{CO}\,\Delta V < 0.056\,\jykms$, or $r_{43}> 0.53$ for
$S_{CO}\,\Delta V < 0.13\,\jykms$.  Both limits on $r_{43}$
suggest \smmN \ has higher excitation, on average, than \smmD,
though whether \smmN \ is more similar to the high-$z$ QSOs and
radio galaxy with CO(1$\rightarrow$0) detections than \smmD \ is
unclear.

To examine the excitation of \smmD \ and \smmN \ in another way,
we plot in Figure~3 the integrated line luminosities for 
different CO lines (hereafter referred to as the CO ladder),
normalized to CO(1$\rightarrow$0), for \smmD \ and \smmN, 
as well as those of the other similarly-luminous high-$z$ galaxies 
with CO(1$\rightarrow$0) detections.  For \smmN, we use our
conservative $3\sigma$ upper limit of $S_{CO}\,\Delta V < 0.13\,\jykms$
in Figure~3. In addition, we include
the CO data of \citet{weiss05}\ for SMM\,J16359+6612 at
$z_{CO} = 2.517$: this galaxy lacks CO(1$\rightarrow$0) data and is an order
of magnitude less luminous than the other high-$z$ sources, 
but it has the most complete CO ladder of any SMG and thus provides
an interesting comparison.  In place of a CO(1$\rightarrow$0) detection
for this galaxy we use the the CO(1$\rightarrow$0) flux implied by 
the LVG fit from \citet{weiss05}.
Also for comparison purposes, we include in Figure~3 
less-IR luminous (by 2-3 orders of magnitude) 
local galaxies commonly used as templates
representative of quiescent galaxies (the Milky Way) and starbursts 
(the center of M82). 
Assuming all the CO emission comes from a single object, 
the CO ladder of \smmD \ appears to rise much more slowly with increasing
$J$ than the ladders of the other high-$z$ active galaxies, 
SMM\,16359+6612, and M82's central region.  The shallow rise of the CO ladder of \smmD \
more resembles the CO ladders of the Milky Way and HR~10, and suggests excitation
conditions in \smmD \ somewhat between those of 
the inner ($2\fdg5 < |l| < 32\fdg5$) and the outer 
($|l| > 32\fdg5$) disk of the Milky Way.
The shape of the CO ladder of the Milky Way (slow rise and turnover at
lower $J$ than the other sources) is caused by significant
sub-thermal excitation, which is any molecular excitation process not governed
by the principles of local thermodynamic equilibrium that causes the excitation
temperature derived by comparing the population of different energy levels
to be lower than the kinetic temperature of the emitting gas.
While we cannot yet say that the CO ladder of \smmD \ turns over in 
a similar way as the Milky Way, the slow rise and low values
of $r_{43}$ suggest that sub-thermal excitation is important in both
\smmD \ and HR~10.  The existence of sub-thermal excitation in these objects
would imply that observations of high-$J$ emission
lines can provide only a lower limit to the molecular gas mass.
\begin{figure}
\includegraphics[width=\columnwidth]{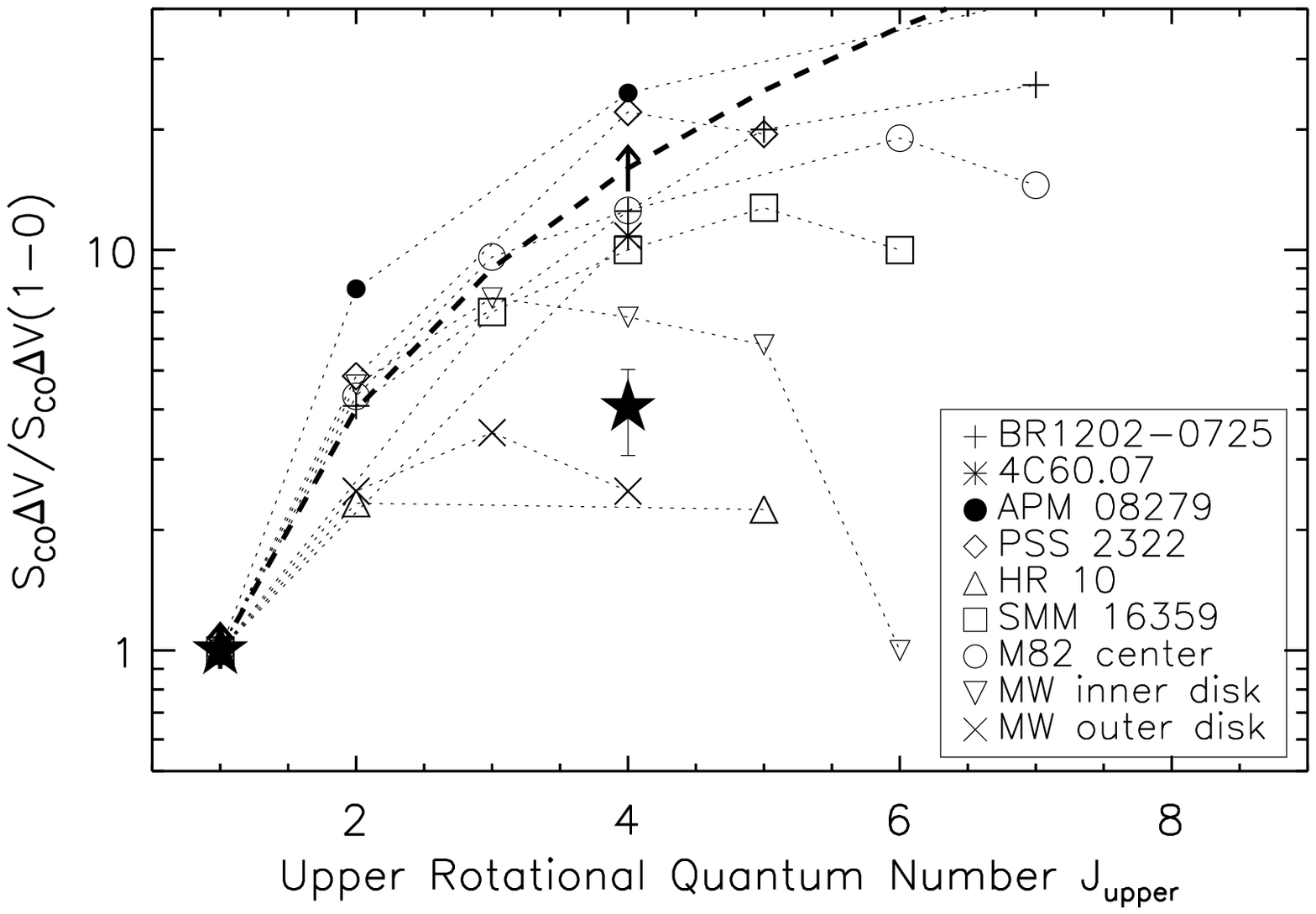}
\caption{CO line ladder for 
high-$z$ galaxies and local templates, normalized to the 
integrated CO(1$\rightarrow$0) flux. The data points for \smmD \ 
and \smmN \ are the filled stars and the upward arrows, respectively.
The thick dashed line represents the effect of constant brightness
temperature on the CO line ratios.   
The CO transition data sources for BR~1202-0725 are 
\citet{omont96}\ and \citet{carilli02}; for 4C\,60.07 \citet{thomasRG}\ 
and \citet{papadop00}; for APM~08279+5255 \citet{downes99}\ and \citet{apm_co10}; 
for PSS~2322+1944 \citet{pss_co10}\ and \citet{cox02};
for HR~10 \citet{andreani00}\ and \citet{HR10co10};
for SMM~16359+6612 \citet{weiss05}\ and \citet{kneib05};
for M82 \citet{mao00}\ and \citet{weiss05a}; and for the Milky Way
\citet{fixsen99}. }
\end{figure}

Our findings for \smmD, along with the results of \citet{HR10co10}\
for HR~10, highlight the need for observations of low-$J$ 
CO emission to measure the entire molecular gas reservoir in high-$z$ galaxies.  
Even though it is true for some high-$z$ galaxies that 
$J\geq3$ CO lines can trace low-excitation gas molecules 
as well as the highly excited gas, such a conclusion 
seems not to apply to \smmD \ and HR~10.  In addition, some
dispersion clearly exists in the gas properties of galaxies hosting
active nuclei at high redshift, since \smmD \ has both at least a weak
AGN and a significant quantity of molecular gas not 
apparent in observations of high-$J$ CO transitions.

\subsection{Star Formation Timescales and Efficiency}

One of the important issues to address in understanding the role
of SMGs in the large-scale picture of galaxy formation and evolution 
is the duration of star formation.  We can use our new estimate of 
the total molecular gas mass in \smmD \ from CO(1$\rightarrow$0) to  
calculate the length of time the major star formation epoch will last
from its respective star formation rate (SFR), assuming the SFR is constant.  We
define this characteristic star formation timescale as
\begin{equation}
t_{SF} = M(\hh)/SFR.
\end{equation}
We calculate the SFR from the far-IR luminosity $L_{FIR}$ 
using Equation~4 of \citet{omont01},
\begin{equation}
SFR = \delta_{MF} (L_{FIR}/10^{10}\,\lsun)\,\msun\,\textrm{yr}^{-1},
\end{equation}
where $\delta_{MF}$ is a function of the stellar initial mass function (IMF).
We use as $L_{FIR}$ the value of $L_{bol}$ found by \citet{chapman05}\   
for \smmD, multiplied by the correction factor of 0.5  
implied by the results of \citet{kovacs06}, and 
assume a Salpeter IMF to obtain $\delta_{MF}=0.8$.
The resulting SFR and $t_{SF}$ we obtain, and the
$L_{FIR}$ used in the calculation of the quantites, are listed in Table~1.
Assuming all the far-IR emission is a result of star formation,  
we thus obtain a short characteristic star formation timescale
$t_{SF} \sim 200$\,Myr, consistent with the brief burst needed
to form the old spheroids and elliptical galaxies we observe at $z\sim 0$,
and in agreement with the duration of SMG star formation bursts
estimated by \citet{smail04}\ from modeling the 
rest-frame optical broad-band colors of SMGs (100-200\,Myr).
Even though the value of $t_{SF}$ we find for \smmD \ is short, it is
significantly longer than 
the star formation timescale calculated from the mass implied by the galaxy's 
CO(4$\rightarrow$3) emission and the assumption of a constant
brightness temperature, which is only 50\,Myr.  The 
CO(1$\rightarrow$0) emission clearly indicates that there is more cold
gas available in \smmD \ to fuel star formation than is
suggested by observations of CO(4$\rightarrow$3) and the assumption of
$r_{43} = 1$.

The continuum-to-line ratio $L_{FIR}/L^{\prime}_{CO}$ provides another 
means of investigating the evolutionary progression of \smmD,
because it is commonly used 
to estimate the efficiency with which molecular gas is converted 
to stars, independent of the assumed value of the 
CO-to-$\textrm{H}_{2}$ conversion factor.
For \smmD, we calculate $L_{FIR}/L^{\prime}_{CO}$ using again 
the value of $L_{bol}$ from \citet{chapman05}, corrected 
as suggested by \citet{kovacs06}, for $L_{FIR}$, 
and we assume all of the CO(1$\rightarrow$0)
emission comes from a single object.  If we assume all the
far-IR emission is a consequence of star formation we obtain an upper limit to the
$L_{FIR}/L^{\prime}_{CO}$ ratio of $\sim 51\,\lsun\,(\Kkmspc)^{-1}$,
significantly less than the value ($\sim 190\,\lsun\,(\Kkmspc)^{-1}$)
obtained with the $J=4\rightarrow3$ line.  

In Table~1 we also list limits on $t_{SF}$ and $L_{FIR}/L^{\prime}_{CO}$ 
for \smmN.  We use the value of $L_{FIR}$ from \citet{neri03}\ with 
our CO(1$\rightarrow$0) upper limits to estimate the continuum-to-line 
ratio, finding $t_{SF} < 15$\,Myr and
$L_{FIR}/L^{\prime}_{CO} > 680\,\lsun\,(\Kkmspc)^{-1}$ for the statistical
upper limit $S_{CO}\,\Delta V < 0.056\,\jykms$, and $t_{SF} < 34$\,Myr and 
$L_{FIR}/L^{\prime}_{CO} > 290\,\lsun\,(\Kkmspc)^{-1}$ for
the more conservative limit $S_{CO}\,\Delta V < 0.13\,\jykms$.  The values 
found from our conservative limit are consistent with the results
from CO(4$\rightarrow$3).

Local LIRGs and ULIRGs are known to trace a scaling relation between
$L^{\prime}_{CO}$ and $L_{FIR}$ in which galaxies with higher IR
luminosities have higher CO luminosities \citep[e.g.][]{young86}.
\citet{yao03}\ find a median value of $L_{FIR}/L^{\prime}_{CO}$
for LIRGs of $50\pm30\,\lsun\,(\Kkmspc)^{-1}$,
while \citet{solomon97}\ obtain a median value for ULIRGs of
$160\pm130\,\lsun\,(\Kkmspc)^{-1}$.   
The high-redshift QSOs and radio galaxies with CO(1$\rightarrow$0) 
detections have continuum-to-line ratios
$L_{FIR}/L^{\prime}_{CO} \sim 200-300\,\lsun\,(\Kkmspc)^{-1}$
from both the CO(1$\rightarrow$0) and CO(4$\rightarrow$3) lines, which is 
larger than that of \smmD \ but comparable to \smmN. 
However, in HR~10, $L_{FIR}/L^{\prime}_{CO}=130\,\lsun\,(\Kkmspc)^{-1}$
from CO(1$\rightarrow$0), nearly an order of magnitude smaller
than the continuum-to-line ratio calculated from its
CO(5$\rightarrow$4) line. Thus, the continuum-to-line ratio $L_{FIR}/L^{\prime}_{CO}$
in \smmD \ resembles local star-forming galaxies and HR~10, 
suggesting similar star formation efficiencies among them, whereas \smmN \ 
appears more like the other high-$z$ systems observed in CO(1$\rightarrow$0).

It thus seems that the physical conditions in the interstellar
medium of \smmD, while similar to another SMG (HR~10), are different than
the high-$z$ QSOs, 4C\,60.07, and \smmN.
Several possibilities could explain the different conditions.
It is possible that the QSOs and 4C\,60.07 truly
have higher star-formation efficiencies than HR~10 and \smmD.
Also, the optically- or radio-luminous high-$z$ galaxies and \smmN \ could 
have more significant contributions to $L_{FIR}$ from dust heated by their active
nuclei, causing an overestimation of the star formation efficiency when 
calculated from the continuum-to-line ratio.  
However, it should be considered that the $L_{FIR}$--$L^{\prime}_{CO}$
relation appears to be non-linear at high values of $L_{FIR}$. 
\citet{gao04a,gao04b}\ suggest that, for this reason, the ratio
$L_{FIR}/L^{\prime}_{CO}$ is not a good indicator of the efficiency of star formation in
the most far-IR-bright galaxies such as high-$z$ QSOs and SMGs.
They argue that HCN emission is a better tracer of the dense gas
associated with star formation and IR emission than CO due to its higher critical
density for excitation ($\sim 10^{5}\,\textrm{cm}^{-3}$), and
since they find $L_{FIR}$ and $L^{\prime}_{HCN}$ are linearly
correlated over three orders of magnitude in $L_{FIR}$, they
suggest that the ratio $L_{FIR}/L^{\prime}_{HCN}$ may be a better indicator 
of star formation efficiency than $L_{FIR}/L^{\prime}_{CO}$.  
In this light, it may be more appropriate to use $L_{FIR}/L^{\prime}_{HCN}$ ratios
or even $L_{FIR}/L^{\prime}_{CO}$ ratios derived from $J\geq 3$
CO transitions, which trace higher-density gas than CO(1$\rightarrow$0), 
to compare the star formation efficiencies of the similarly luminous
PSS~2322+1944, APM~08279+5255, and 4C\,60.07 to \smmD \ and HR~10.

\subsection{A Merger in \smmD?}

In \S4.1, we found the velocity width of the Gaussian fit to
the CO(1$\rightarrow$0) line profile of \smmD \ to be
$\Delta V_{FWHM}=1040\pm190\,\kms$, which deviates significantly
from the FWHM fitted to the CO(4$\rightarrow$3) line by \citet{grevesmg},
$530\pm50\,\kms$.  Such a discrepancy is not typically expected; in general,
we expect the line profiles of CO transitions of different $J$
will be of similar widths, from observations 
of nearby galaxies \citep[e.g.][]{yao03}.
However, as mentioned in \S4.1, our data are not of sufficient
quality to be certain that a Gaussian profile is an appropriate
model for the line shape, so it may be misleading to compare
the Gaussian FWHM velocity widths.  To determine the reality of
the factor of $\sim 2$ difference in velocity width 
from the fits, we instead directly compare 
the CO(1$\rightarrow$0) spectrum of \smmD \ with the CO(4$\rightarrow$3)
spectrum from \citet{grevesmg}\ in Figure~4, where the amplitude of 
the 1$\rightarrow$0 spectrum has been multiplied 
by a factor of 5 in both panels for ease of comparison.  In panel (a), 
we compare the spectra smoothed to the same velocity resolution shown
in \citet{grevesmg}, $58\,\kms$, and it is apparent that
the low S/N ratio of both spectra preclude
detailed comparison.  Thus, in panel (b), we further smooth and re-bin
the line profiles of both CO transitions to $170\,\kms$, and find
that the CO(4$\rightarrow$3) line is nearly as wide as the 
CO(1$\rightarrow$0) line, which suggests that the CO(4$\rightarrow$3)
line FWHM measurement may have been underestimated in \citet{grevesmg}.
However, in panel (b) of Figure~4, we observe that the
CO(1$\rightarrow$0) line extends to a lower velocity region of the
spectrum than the CO(4$\rightarrow$3) line, suggesting that
there exists blueshifted (relative to the central system redshift) 
molecular gas in \smmD \ which is less excited than the molecular
gas at more positive velocity offsets.
\begin{figure}
\includegraphics[width=\columnwidth]{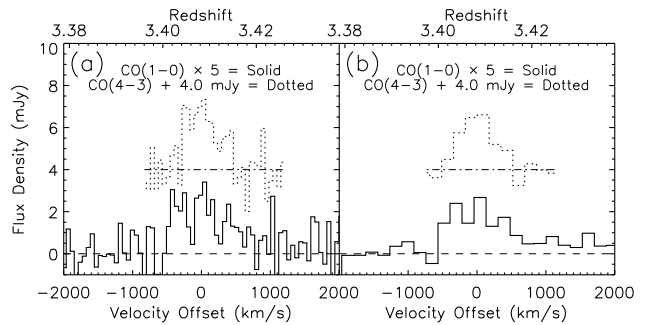}
\caption{Comparison of CO(1$\rightarrow$0) (solid line) and 
CO(4$\rightarrow$3) spectra (dotted line). The CO(4$\rightarrow$3)
spectrum has been provided by T. Greve. 
For clarity, the amplitude of the 
CO(1$\rightarrow$0) spectrum has been multiplied by a factor of 5
and the CO(4$\rightarrow$3) spectrum has been offset from zero by 4\,mJy.
The spectra are smoothed to a common velocity resolution of 58\,$\kms$ in panel (a) 
and $170\,\kms$ in panel (b).}
\end{figure}

Possible explanations of this excitation variation with velocity arise
from both of the dynamical scenarios discussed by \citet{tacconi06}\
to explain the double-peaked CO line profiles they observe from
other SMGs: a rotating disk and a merger of several galaxy components.
If \smmD \ were a single rotating disk, the blueward
extension of the CO(1$\rightarrow$0) line profile could indicate
a physically distinct region of cold gas within the disk not
symmetrically distributed about the more
highly excited region represented by the CO(4$\rightarrow$3) emission,
such as a single giant molecular cloud complex.
On the other hand, in a rotating disk the rotation curve
should flatten away from the center of the galaxy.  In such a case, when 
the CO line profiles are observed averaged over the entire galaxy,
the width of the CO(1$\rightarrow$0) profile would not increase 
over the CO(4$\rightarrow$3) profile even if the lower excitation
gas comes from an extended region as compared with the location of the
CO(4$\rightarrow$3) emission.  Alternatively, if \smmD \ is 
an interaction or a merger of several subcomponents separated by several hundred
$\kms$ in velocity space, the CO(1$\rightarrow$0) line profile
could be different from the CO(4$\rightarrow$3) profile if the merging
components possess different excitation conditions.

Continuum observations of \smmD \ at different wavelengths seem to
support a scenario involving an interaction or merger of 
galaxy components.  In Figure~5 we show previously unpublished  
optical, near-infrared, and radio continuum images of \smmD,
from Borys et al.\ (2006, in preparation) and Fomalont et al.\ 
(2006, private communication).
\begin{figure*}[tb]
\centering
\includegraphics*[width=2.0\columnwidth]{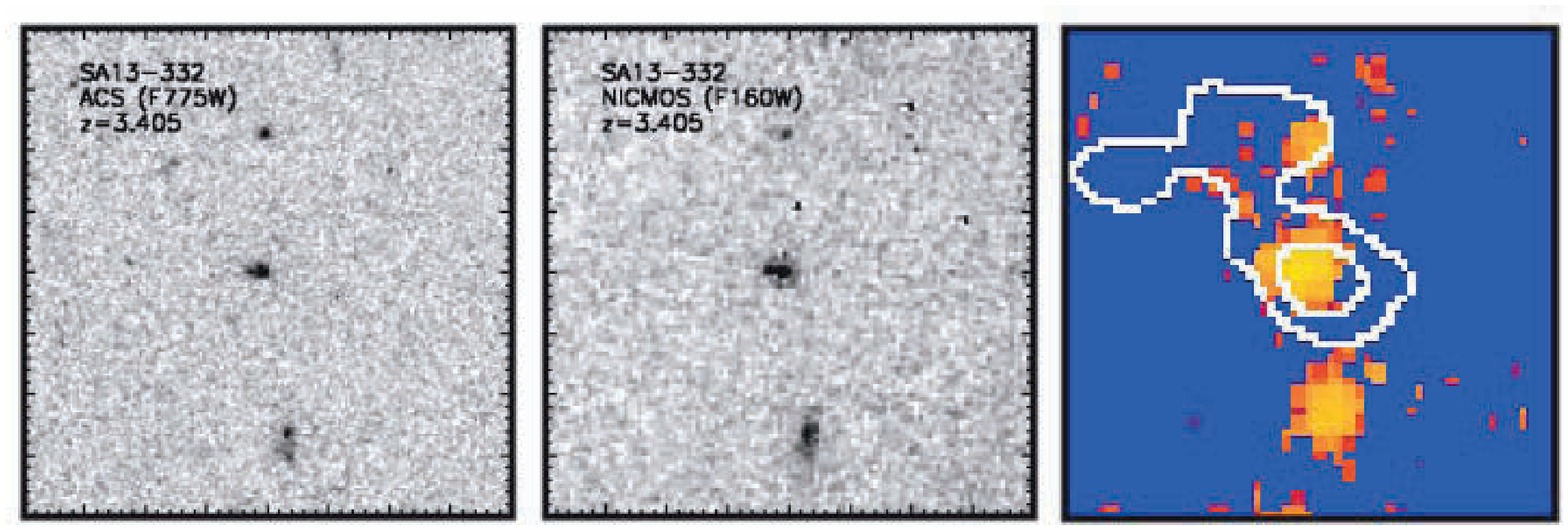}
\caption{$8\arcsec \times 8\arcsec$ images centered on the radio 
position of \smmD \ ($\alpha_{J2000}=13^{h}12^{m}01\fs172$, 
$\delta_{J2000}=42\degr42\arcmin08\farcs39$). 
\emph{Left Panel}: $8\arcsec \times 8\arcsec$ 
HST-ACS F775W-band ($i^{\prime}$) image (Borys et al.\ 2006, 
in preparation). \emph{Center Panel}: HST-NICMOS F160W-band image (Borys et al.\ 
2006, in preparation). \emph{Right Panel}: VLA 1.4\,GHz contours plotted
over a Subaru-SUPRIMECam $R$-band image (Fomalont et al.\ 2006, private communication).
The VLA data were obtained in A-array, with synthesized beam size $1\farcs 5$, and
the contour levels in the image are 15 and $30\,\mu\textrm{Jy}$.}
\end{figure*}
At the position of the 1.4\,GHz emission peak the HST
F775W- and F160W-band images show a compact, close-separation 
($\sim 0\farcs2$) double source.  Plus, the radio 
continuum emission appears to be extended on the scale of
$\sim 3\arcsec$ (at P.A. $\simeq 42\degr$), significant at
the $4\sigma$ level.  In fact, the northern source 
$\sim 2\farcs5$ away in the HST images appears to be 
associated with this extended radio emission.  The radio
continuum indicates significant activity (from black hole accretion
or star formation) outside the radio/optical core of \smmD, 
and is suggestive of a galaxy interaction or merger.  However, it is not 
necessarily associated with the blueward-extending CO(1$\rightarrow$0),
since the activity causing the extended radio emission would likely 
cause increased CO excitation.  Obtaining high-accuracy redshifts
of the optical sources around the optical/radio core 
and/or interferometric imaging at high spatial
resolution of the CO emission lines can
verify the merger/interaction scenario: if the optical/radio sources 
are found to have redshifts corresponding to the center and outer
regions of the CO(1$\rightarrow$0) and CO(4$\rightarrow$3) lines, then the
sources may indeed be interacting galaxies.

\lastpagefootnotes
More data are clearly needed to determine the 
structure and energetics
of the major gas component(s) in \smmD, and SMGs as a class.
In particular, spatially- and spectrally-resolved
images and spectra of CO(1$\rightarrow$0) and high-$J$ emission
will be the most revealing, along with HCN observations, and new
instruments and facilities planned for the future will
help provide these. In the near future, the new Ka-band correlation
receiver at GBT will cover the frequency range to which CO(1$\rightarrow$0) is 
redshifted for galaxies in the range $2<z<3$ (26--40\,GHz) and thus will aid
the detection of CO(1$\rightarrow$0) in many more SMGs in a modest
integration time ($\sim 6$\,hours or less).  The GBT will also be important
in obtaining HCN data for \smmD \ and SMGs in general, 
in a few years assisted by the EVLA\footnote{Information on the
EVLA project is available at http://www.aoc.nrao.edu/evla/.}.  
Looking further in the future, ALMA\footnote{Information on ALMA
may be obtained from http://www.alma.nrao.edu/.} will be 
instrumental in understanding the nature of \smmD \ and other SMGs, 
since it will provide spatial resolution comparable to HST at 
submillimeter wavelengths and allow us to determine if and 
where there are multiple components contributing to the far-IR continuum emission.
ALMA will also enable high-quality imaging of many 
CO and HCN transitions in SMGs, so that the spatial dependence
of excitation conditions can be explored.

\acknowledgments

We gratefully acknowledge the assistance and advice 
of R. Maddalena and A. Minter at NRAO-Green Bank in 
developing tools to calibrate the GBT K-band data.  We also thank
C. Borys for providing HST images of \smmD \ and E. Fomalont for providing VLA
and Subaru images of \smmD.  We wish to thank the anonymous referee for 
helpful comments which improved the manuscript.  LJH
acknowledges the support of the GBT Graduate Funding program during this work.
AWB acknowledges support from the Alfred P. Sloan Foundation and the Research Corporation.
IRS acknowledges support from the Royal Society.
The Green Bank Telescope is a facility of the National Radio Astronomy Observatory, 
operated by Associated Universities, Inc., under a cooperative 
agreement with the National Science Foundation.

\clearpage

\begin{deluxetable*}{lcccccccccc}[b]
\tabletypesize{\tiny}
\tablewidth{0pt}
\tablenum{1}
\tablecolumns{11}
\tablecaption{Physical Parameters of Observed SMGs}
\tablehead{
\colhead{Source} & \colhead{$z_{CO}$} & \colhead{$\mu_{L}$\tablenotemark{a}} & 
\colhead{$L_{FIR}$} & \colhead{$SFR$\tablenotemark{b}} & \colhead{$r_{43}$} & \colhead{CO Trans} &
\colhead{$M(\hh)$} & \colhead{$t_{SF}$\tablenotemark{c}} & \colhead{$L_{FIR}/L^{\prime}_{CO}$} & 
\colhead{Reference} \\
\colhead{} & \colhead{} & \colhead{} & \colhead{($10^{13}\,\lsun$)} & 
\colhead{($\msun\,\textrm{yr}^{-1}$)} & \colhead{} & \colhead{} & \colhead{($10^{10}\,\msun$)} &
\colhead{(Myr)} & \colhead{($\lsun\,(\Kkmspc)^{-1}$)} & \colhead{}
}
\startdata
\smmD & 3.408 & 1.0 & 1.0 & 810 & $0.26\pm0.06$ & 1$\rightarrow$0 & $16\pm3$ & 200 & 51 & (1),(2),(3) \\
    &  &  &  &  &  &                                        4$\rightarrow$3 & $4.2\pm0.7$ & 50 & 190 &     \\
\smmN \tablenotemark{d} & 3.346 & 1.2 & 1.5 & 1200 & $> 0.53$ & 1$\rightarrow$0 & $< 4.1$ & $< 34$ & $> 290$ & (1),(4) \\  
    &  &  &  &  &  &                                        4$\rightarrow$3 & $2.1\pm0.2$ & 18 & 560 &     \\

\enddata
\tablenotetext{a} {Gravitational lensing magnification factor applied to $L_{FIR}$ and $L^{\prime}_{CO}$, and thus
included in all quantities calculated from those luminosities.}
\tablenotetext{b}{Derived from $L_{FIR}$ according to Equation~7 in the text.}
\tablenotetext{c}{Calculated using Equation~6 in the text.}
\tablenotetext{d}{``Conservative'' CO(1$\rightarrow$0) limits from \S4.2 used in all calculations.}
\tablerefs{(1) Greve et al.\ 2005; (2) Chapman et al.\ (2005); (3) Kov\'{a}cs et al.\ 2006; (4) Neri et al.\ 2003}
\end{deluxetable*}

\end{document}